\documentclass[twocolumn]{aastex63}

\received{}
\revised{}
\accepted{}

\shortauthors{Yang et al.}

\begin{document}

\turnoffeditone
\newcommand{\kms}{km\,s$^{-1}$}
\newcommand{\cii}{[\ion{C}{2}]}
\newcommand{\ci}{[\ion{C}{1}]}
\newcommand{\oi}{[\ion{O}{1}]}
\newcommand{\mgii}{\ion{Mg}{2}}
\newcommand{\feii}{\ion{Fe}{2}}
\newcommand{\civ}{\ion{C}{4}}

\title{P\={o}niu\={a}`ena: A Luminous $z=7.5$ Quasar Hosting a 1.5 Billion Solar Mass Black Hole}

\correspondingauthor{Jinyi Yang}
\email{jinyiyang@email.arizona.edu}

\author[0000-0001-5287-4242]{Jinyi Yang}
\affil{Steward Observatory, University of Arizona, 933 N Cherry Ave, Tucson, AZ 85721, USA}

\author[0000-0002-7633-431X]{Feige Wang}
\altaffiliation{NHFP Hubble Fellow}
\affil{Steward Observatory, University of Arizona, 933 N Cherry Ave, Tucson, AZ 85721, USA}

\author[0000-0003-3310-0131]{Xiaohui Fan}
\affil{Steward Observatory, University of Arizona, 933 N Cherry Ave, Tucson, AZ 85721, USA}

\author[0000-0002-7054-4332]{Joseph F. Hennawi}
\affil{Department of Physics, University of California, Santa Barbara, CA 93106-9530, USA}
\affil{Max Planck Institut f\"ur Astronomie, K\"onigstuhl 17, D-69117, Heidelberg, Germany}

\author[0000-0003-0821-3644]{Frederick B. Davies}
\affil{Department of Physics, University of California, Santa Barbara, CA 93106-9530, USA}
\affil{Lawrence Berkeley National Laboratory, 1 Cyclotron Rd, Berkeley, CA 94720, USA}

\author[0000-0002-5367-8021]{Minghao Yue}
\affil{Steward Observatory, University of Arizona, 933 N Cherry Ave, Tucson, AZ 85721, USA}

\author[0000-0002-2931-7824]{Eduardo Banados}
\affil{Max Planck Institut f\"ur Astronomie, K\"onigstuhl 17, D-69117, Heidelberg, Germany}

\author[0000-0002-7350-6913]{Xue-Bing Wu}
\affil{Kavli Institute for Astronomy and Astrophysics, Peking University, Beijing 100871, China}
\affil{Department of Astronomy, School of Physics, Peking University, Beijing 100871, China}

\author[0000-0001-9024-8322]{Bram Venemans}
\affil{Max Planck Institut f\"ur Astronomie, K\"onigstuhl 17, D-69117, Heidelberg, Germany}

\author[0000-0002-3026-0562]{Aaron J. Barth}
\affil{Department of Physics and Astronomy, University of California, Irvine, CA 92697, USA}

\author[0000-0002-1620-0897]{Fuyan Bian}
\affil{European Southern Observatory, Alonso de C\'ordova 3107, Casilla 19001, Vitacura, Santiago 19, Chile}

\author[0000-0003-4432-5037]{Konstantina Boutsia}
\affil{Las Campanas Observatory, Carnegie Observatories, Colina El Pino, Casilla 601, La Serena, Chile}

\author[0000-0002-2662-8803]{Roberto Decarli}
\affil{INAF -- Osservatorio di Astrofisica e Scienza dello Spazio di Bologna, via Gobetti 93/3, I-40129 Bologna, Italy}

\author[0000-0002-6822-2254]{Emanuele Paolo Farina}
\affiliation{Max Planck Institut f\"ur Astrophysik, Karl--Schwarzschild--Stra{\ss}e 1, D-85748, Garching bei M\"unchen, Germany}

\author[0000-0003-1245-5232]{Richard Green}
\affil{Steward Observatory, University of Arizona, 933 N Cherry Ave, Tucson, AZ 85721, USA}

\author[0000-0003-4176-6486]{Linhua Jiang}
\affil{Kavli Institute for Astronomy and Astrophysics, Peking University, Beijing 100871, China}

\author[0000-0001-6239-3821]{Jiang-Tao Li}
\affil{Department of Astronomy, University of Michigan, 311 West Hall, 1085 S. University Ave, Ann Arbor, MI, 48109-1107, USA}

\author[0000-0002-5941-5214]{Chiara Mazzucchelli}
\affil{European Southern Observatory, Alonso de C\'ordova 3107, Casilla 19001, Vitacura, Santiago 19, Chile}

\author[0000-0003-4793-7880]{Fabian Walter}
\affil{Max Planck Institut f\"ur Astronomie, K\"onigstuhl 17, D-69117, Heidelberg, Germany}



\begin{abstract}
  We report the discovery of a luminous quasar, J1007+2115 at $z=7.515$ (``P\={o}niu\={a}`ena''), from our wide-field reionization-era quasar survey. J1007+2115 is the second quasar now known at $z>7.5$, deep into the reionization epoch.
The quasar is powered by a $(1.5\pm0.2)\times10^9$\,$M_{\odot}$ supermassive black hole (SMBH), based on its broad \ion{Mg}{2} emission-line profile from Gemini and Keck near-IR spectroscopy.  The SMBH in J1007+2115 is twice as massive as that in quasar J1342+0928 at $z=7.54$, the current quasar redshift record holder. 
The existence of such a massive SMBH just 700 million years after the Big Bang significantly challenges models of the earliest SMBH growth. Model assumptions of
Eddington-limited accretion and a radiative efficiency of 0.1 require a seed black hole of $\gtrsim 10^{4}$\,$M_{\odot}$ at $z=30$.
This requirement suggests either a massive black hole seed as a result of direct collapse or earlier periods of rapid black hole growth with hyper-Eddington accretion and/or a low radiative efficiency.
 We measure the damping wing signature imprinted by neutral hydrogen absorption in the intergalactic medium (IGM) on J1007+2115's Ly$\alpha$ line profile, and find that it is weaker than that of J1342+0928 and two other $z\gtrsim7$ quasars. We estimate an IGM volume-averaged neutral fraction $\langle x\rm_{HI}\rangle=0.39^{+0.22}_{-0.13}$. This range of values suggests a patchy reionization history toward different IGM sightlines.
We detect the 158 $\mu$m \cii\ emission line in J1007+2115 with ALMA; this line centroid yields a systemic redshift of $z=7.5149\pm0.0004$ and indicates a star formation rate of $\sim210$\,$M_\sun$\,yr$^{-1}$ in its host galaxy. 

\end{abstract}

\keywords{galaxies: active --- galaxies: high-redshift --- quasars: individual (UHS J100758.264+211529.207) --- cosmology: observations --- early universe }


\section{Introduction} 
Luminous reionization-era quasars ($z> 6.5$) provide unique probes of supermassive black hole (SMBH) growth, massive galaxy formation, and intergalactic medium (IGM) evolution in the first billion years of the Universe's history. However, efforts to find such objects have proven to be difficult because of a combination of the declining spatial density of quasars at high redshift, the limited sky coverage of near-infrared (NIR) photometry, and the low efficiency of spectroscopic follow-up observations.

During the past few years, high-redshift quasar searches using newly available wide-area optical and IR surveys have resulted in a sixfold increase in the number of known quasars at $z>6.5$:
47 luminous quasars at $z>6.5$ have been discovered \citep[e.g.,][]{fan19,venemans13,venemans15,wang19,mazzucchelli17,reed19,matsuoka19a,yang19}, although among them only six are at $z\ge7$ \citep{mortlock11, wang18, matsuoka19a, matsuoka19b, yang19} and one at $z>7.1$ \citep{banados18}.
These discoveries show that 800 million solar-mass black holes exist already at $z=7.5$ \citep{banados18} and that the IGM is significantly neutral at $z\gtrsim7$ \citep[e.g.][]{banados18,davies18b,greig17,greig19,wang20}.
However,  both early SMBH growth history and IGM neutral fraction evolution at $z>7$ are still poorly constrained because of the small sample size. 
For statistical analysis, more $z\sim7-8$ quasars are necessary to investigate the IGM, SMBH masses, and quasar host galaxies at this critical epoch.

In this paper, we report the discovery of a new quasar J100758.264+211529.207 (``P\={o}niu\={a}`ena'' in the Hawaiian language, hereinafter J1007+2115) at $z=7.5149$. This object is only the second quasar known at $z \sim 7.5$, close to the mid-point redshift of reionization \citep{planck18}. Its discovery enables new measurements of a quasar Ly$\alpha$ damping wing and provides new constraints on the earliest SMBH growth. In this paper, we adopt a $\Lambda$CDM cosmology with parameters $\Omega_{\Lambda}=0.7$, $\Omega_{m}=0.3$, and $h=0.685$. Photometric data are reported on the AB system after applying a Galactic extinction correction \citep{schlegel98,schlafly11}. 

\section{Candidate Selection and Observations}
In this section we describe the selection method that led to the discovery of J1007+2115 and the spectroscopic observations. This quasar was selected based on the same photometric dataset used for our previous $z\sim6.5-7$ quasar surveys \citep{wang18,wang19,yang19} but with selection criteria focused on a higher redshift range.

\subsection{Selection Method}
We have constructed an imaging dataset by combining all available optical and infrared photometric surveys that covers $\sim$ 20,000 deg$^{2}$ of high Galactic latitude sky area with $z/y, J$, and WISE photometry to the depth of $J \sim 21$ ($5\sigma$), and have used this dataset to carry out a wide-field systematic survey for quasars at $z>6.5$ \citep{wang18,wang19,yang19}.
J1007+2115 was selected in the area covered by the DESI Legacy Imaging Surveys \citep[DECaLS,][]{dey19}, the Pan-STARRS1 \citep[PS1,][]{chambers16} survey, the UKIRT Hemisphere Survey \citep[UHS,][]{dye18}, and the {\em Wide-field Infrared Survey Explorer} survey \citep[\emph{WISE,}][]{wright10}.
For the {\em WISE} photometry, when we applied the selection cuts, we used the photometric data from the ALLWISE catalog \footnote{\url{http://wise2.ipac.caltech.edu/docs/release/allwise/}}.
To identify quasars at $z\gtrsim7.5$, we required the object to be undetected in all optical bands.
We used a simple IR color cut $J - W1 > -0.261$, S/N$_{J}$ $>$ 5, S/N$_{W1}$ $>$ 5. Forced aperture photometry in all PS1 and DECaLS bands was used to reject contaminants further. After the selection cuts, we visually inspected images of each candidate in all bands. In this step, both the ALLWISE and unWISE \citep{lang14,meisner18} images were included. All photometric data are summarized in Table \ref{tab:table1}.

\subsection{Near-infrared Spectroscopy}
\begin{figure*}
\centering 
\epsscale{0.865}
\plotone{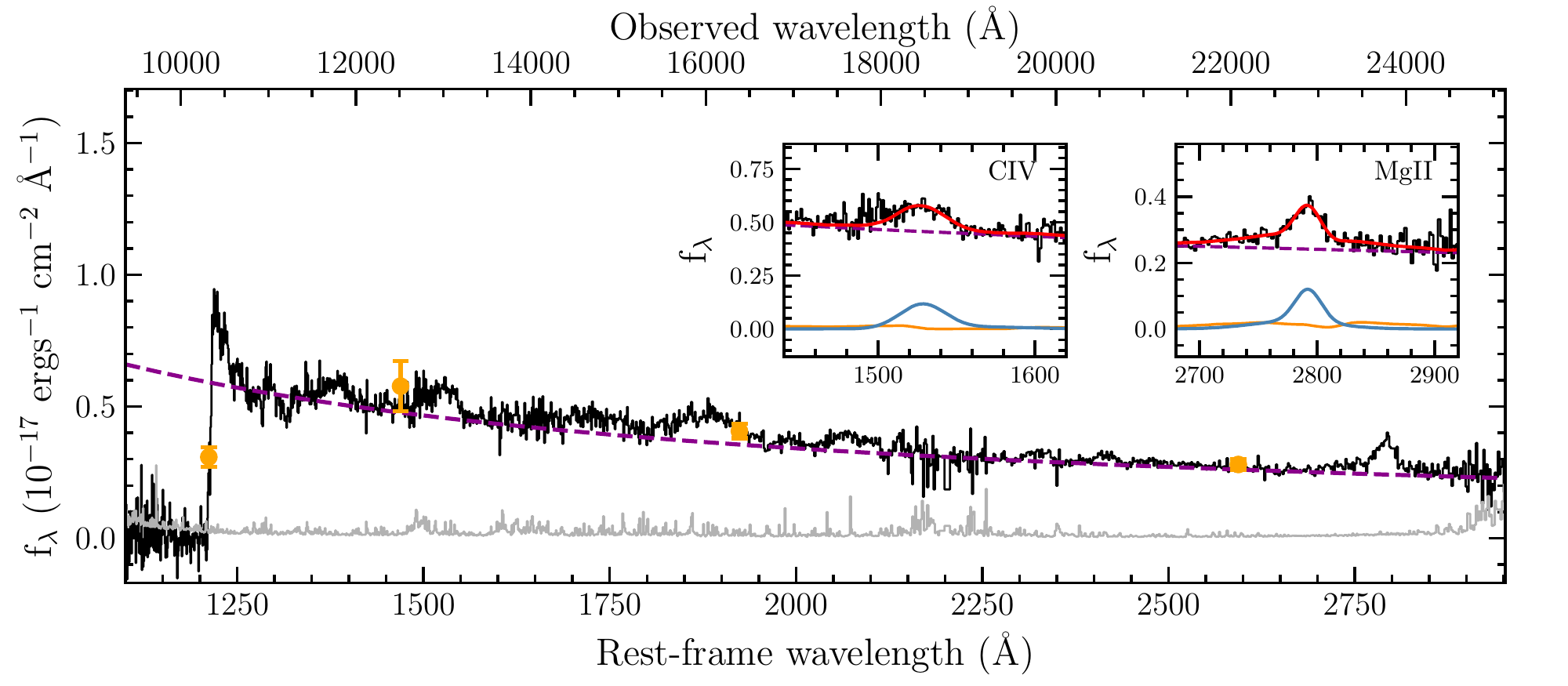} 
\hspace{-15pt}
\epsscale{0.30}
\plotone{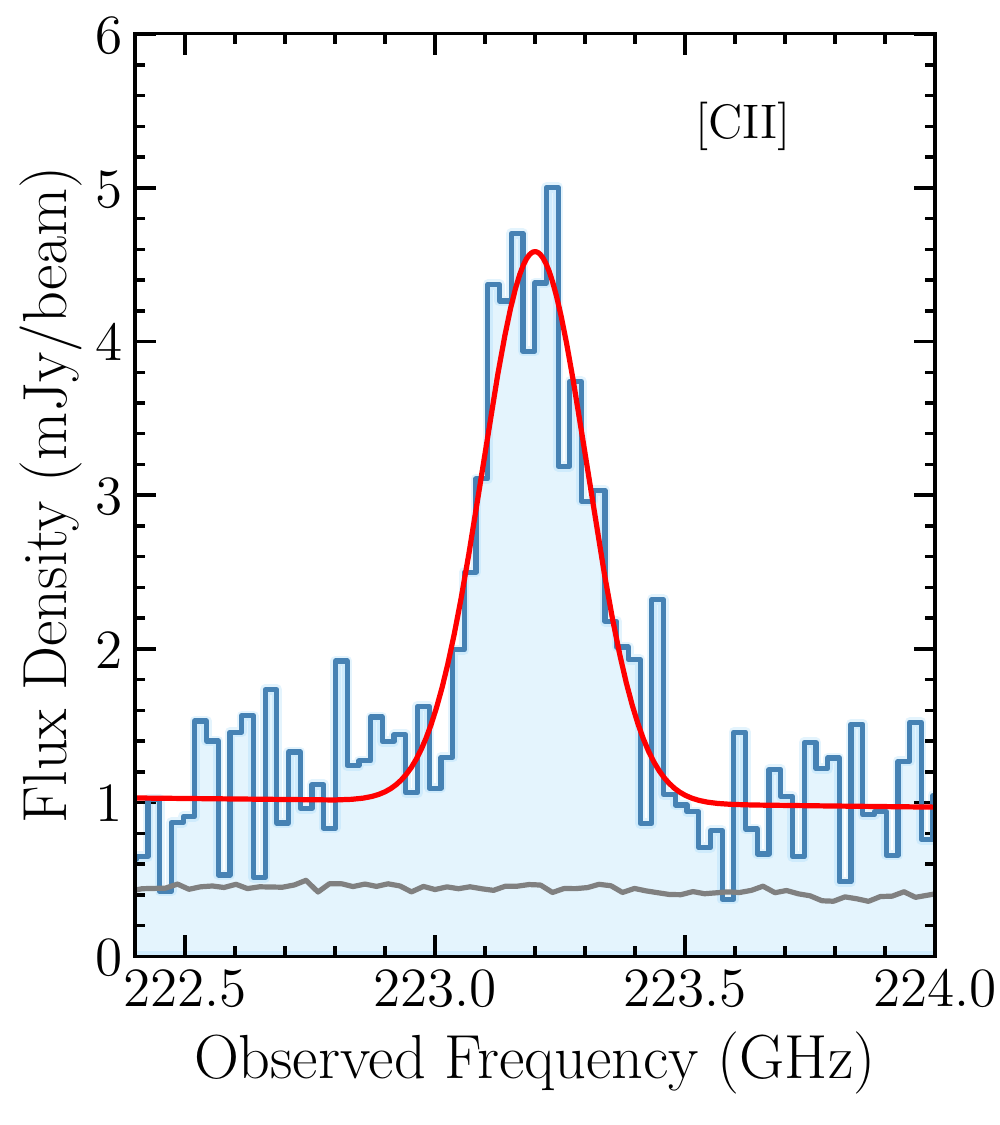} 
\epsscale{1.2}
\plotone{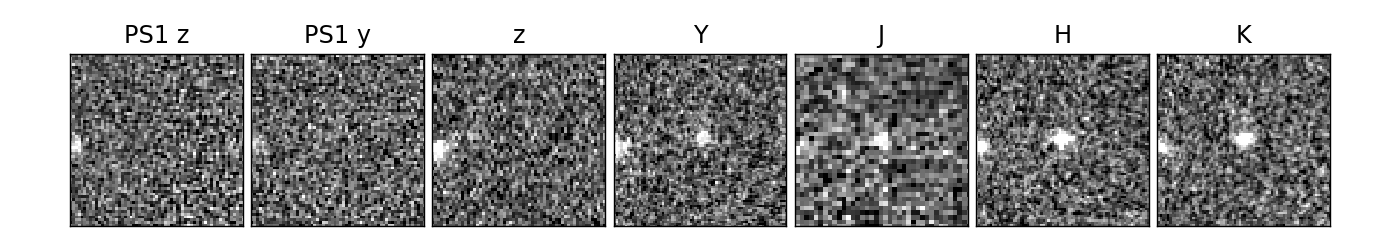} 
\caption{{\bf Upper-left}: The combined spectrum of J1007+2115 from GNIRS and NIRES data, compared with photometric data in the $Y,J,H$, and $K$ bands (orange points with error bars). The $J$-band data point is from the UHS and data in other three bands are from our photometry with UKIRT obtained after the discovery of this quasar. The photometric data are consistent with the spectrum. The purple dashed line represents the best-fit pseudo-continuum. The two inner plots show the fits to the \civ\ and \mgii\ lines. The red solid lines represent the best-fit spectra. The orange lines are the \feii\ components and the blue lines denote the best-fit emission lines. {\bf Upper-right}: The spectrum of the \cii\ emission line with the uncertainty (grey) and best fit Gaussian profile (red). The \cii\ line peaks at the observed frequency 223.2$\pm$0.01 GHz, corresponding to a redshift of 7.5149$\pm$0.0004. {\bf Bottem}: Images ($15'' \times 15''$, north is up and east is to the left) of J1007+2115 in PS1 $z$, PS1 $y$, DECaLS $z$, UKIRT $Y$, UHS $J$, UKIRT $H$, and UKIRT $K$ bands. This quasar is not detected in PS1 $z$, PS1 $y$, and DECaLS $z$. The 3$\sigma$ flux limits in these three bands are measured from our forced photometry and reported in Table \ref{tab:table1}.}
\label{fig:spec}
\end{figure*}

J1007+2115 was confirmed as a quasar during our Gemini/GNIRS run in 2019 May. The discovery spectrum was of low quality because of the high airmass when it was observed. A one-hour (on-source) observation with Magellan/FIRE was used further to confirm this new quasar shortly after the GNIRS observations. To obtain higher quality spectra, we observed the quasar for 5.5 hours (on-source) with GNIRS and for 2.2 hours (on-source) with Keck/NIRES in May and June of 2019. The redshift measured from the \mgii\ line is $z_{\rm MgII}=7.494\pm0.001$. Since J1007+2115 was first discovered with the Gemini North Telescope in Hawaii, J1007+2115 was given the Hawaiian name ``P\={o}niu\={a}`ena'', which means ``unseen spinning source of creation, surrounded with brilliance" in the Hawaiian language.

With Gemini/GNIRS, we used the short-slit (cross-dispersion) mode (32 l/mm) with simultaneous coverage of 0.85--2.5 $\mu$m. A 1$\farcs$0 slit ($R\sim400$) was used for the discovery observations, while a 0$\farcs$675 slit ($R\sim620$) was used for the additional high quality spectrum. For Magellan/FIRE, the echelle mode with a 0$\farcs$75 slit ($R\sim4800$) was used. Keck/NIRES has a fixed configuration that simultaneously covers 0.94 to 2.45 $\mu$m with a fixed $0\farcs55$ narrow slit resulting in a resolving power of $R\sim2700$ \citep{wilson04}. 
All NIR spectra were reduced with the  open-source Python-based spectroscopic data reduction pipeline {\tt PypeIt}\footnote{\url{https://github.com/pypeit/PypeIt}} \citep{prochaska20}. We corrected the telluric absorption by fitting an absorption model directly to the quasar spectra using the telluric model grids produced from the Line-By-Line Radiative Transfer Model \citep[{\tt LBLRTM}\footnote{\url{http://rtweb.aer.com/lblrtm.html}};][]{clough05}.
We stacked the spectra from GNIRS and NIRES, weighted by inverse-variance, and scaled the result with the $K$-band magnitude. The final stacked spectrum is shown in Figure \ref{fig:spec}.

\subsection{\cii-based Redshift and Dust Continuum from ALMA}
We observed J1007+2115 with ALMA (configuration C43-4, Cycle 7) to detect the \cii\ emission line and underlying dust continuum emission from the quasar host galaxy. The observations were taken in 2019 October with 15 min on-source integration time. The synthesized beam size is $ 0\farcs46 \times 0\farcs34$ and the final data cube reaches an rms noise level of 0.4 mJy beam$^{-1}$ per 10 km s$^{-1}$ channel. The ALMA data were reduced with the CASA 5.4 pipeline \citep{mcmullin07}. J1007+2115 is strongly detected in both the \cii\ emission line and the continuum. The source is not spatially resolved. 
 
The \cii\ emission line provides the most accurate measurement of the quasar systemic redshift. A Gaussian fit to the \cii\ line yields a redshift of 7.5149$\pm$0.0004. 
We use the \cii-based redshift as the systemic redshift of the quasar. We obtain a line flux of $F_\mathrm{[CII]}=1.2\pm0.1$ Jy\,\kms, and an FWHM$_\mathrm{[CII]}=331.3\pm31.6$ \kms, corresponding to a line luminosity of $L_\mathrm{[CII]}=(1.5\pm0.2)\times10^{9}$\,$L_{\odot}$.
Applying the relation between star formation rate (SFR) and $L_\mathrm{[CII]}$ for high-redshift ($z>0.5$) galaxies from \cite{delooze14} which has a systematic uncertainty of a factor of $\sim$2.5, we obtain SFR$_\mathrm{[CII]}\sim80-520$\,$M_\sun$\,yr$^{-1}$. This is similar to the SFR of quasar J1342+0928 at $z=7.54$ \citep{venemans17}.
The underlying dust continuum is also detected, and we measure 1.2$\pm$0.03 mJy at 231.2 GHz. We obtain far-infrared (FIR: rest-frame 42.5--122.5 $\mu$m) and total infrared luminosities (TIR: 8--1000 $\mu$m) of $L_{\rm FIR}$ = (3.3$\pm$0.1)$\times10^{12}$\,$L_{\odot}$ and $L_{\rm TIR}$ = (4.7$\pm$0.1)$\times10^{12}$\,$L_{\odot}$, assuming an optically thin grey body with dust temperature $T_{d}$ = 47 K and emissivity index $\beta$ = 1.6 \citep{beelen06} and taking the effect of the cosmic microwave background (CMB) on the dust emission into account \citep[e.g.,][]{dacunha13}. The SFR$_{\rm TIR}$ is estimated as $\sim$ 700 $M_{\odot}$/yr by applying the local scaling relation from \cite{murphy11}.

\section{A 1.5 Billion Solar Mass Black Hole}
The central black hole mass of the quasar can be estimated based on its luminosity and the FWHM of the \mgii\ line. 
We fit the near-IR spectrum with a pseudo-continuum, including a power-law continuum, Fe II template \citep{vestergaard01,tsuzuki06}, and Balmer continuum \citep{derosa14}. Gaussian fits of the \civ\ and \mgii\ emission lines are performed on the continuum-subtracted spectrum. A two-component Gaussian profile is used. The uncertainty is estimated using 100 mock spectra created by randomly adding Gaussian noise at each pixel with its scale equal to the spectral error at that pixel \citep[e.g.][]{shen19, wang20}.
All uncertainties are then estimated based on the 16th and 84th percentile of the distribution.
The best-fit pseudo continuum and the line fitting of \civ\ and \mgii\ are shown in Figure \ref{fig:spec}.

From the spectral fit, we find that the power-law continuum has a slope $\alpha=-1.14\pm0.01$ ($f_{\lambda}\propto\lambda^{\alpha}$). The rest-frame 3000 \AA\ luminosity is measured to be $\lambda L_{\rm 3000}=(3.8\pm0.2)\times10^{46}$ erg s$^{-1}$, corresponding to a bolometric luminosity of $L_{\rm bol}=(1.9\pm0.1)\times10^{47}$ erg s$^{-1}$ assuming a bolometric correction factor of 5.15 \citep{richards06}.
The apparent and absolute rest-frame 1450 \AA\ magnitudes are derived to be $m_{\rm 1450,AB}=20.43\pm0.07$ and $M_{\rm 1450,AB}=-26.66\pm0.07$ from the best-fit power law continuum. 
The line fitting of \mgii\ yields an FWHM = $3247\pm188$ km s$^{-1}$ and a \mgii-based redshift of $z_{\rm MgII}=7.494\pm0.001$, implying a $736\pm35$ km s$^{-1}$ blueshift relative to the \cii\ line, similar to other $z\gtrsim7$ luminous quasars \citep[e.g.,][]{mortlock11,banados18,wang20}. The \civ\ fitting results in an FWHM of 6821$\pm$2055 km s$^{-1}$. The \civ\ line has a 3220$\pm$362 km s$^{-1}$ blueshift compared to the \mgii\ line. These measurements are summarized in Table 1.

We estimate the black mass based on the bolometric luminosity and the FWHM of the \mgii\ line by adopting the local empirical relation from \citet{vestergaard09}. The black hole mass is derived to be $M_{\rm BH}=(1.5\pm0.2)\times10^{9}$\,$M_{\odot}$, resulting in an Eddington ratio of $L_{\rm bol}/L_{\rm Edd}=1.1\pm0.2$.
Note that the black hole mass uncertainty estimated here does not include the systematic uncertainties of the scaling relation, which could be up to $\sim0.5$ dex.
The uncertainty of the Eddington ratio is subject to the same systematic uncertainty as the black hole mass. 

\begin{figure}
\centering 
\epsscale{1.25}
\plotone{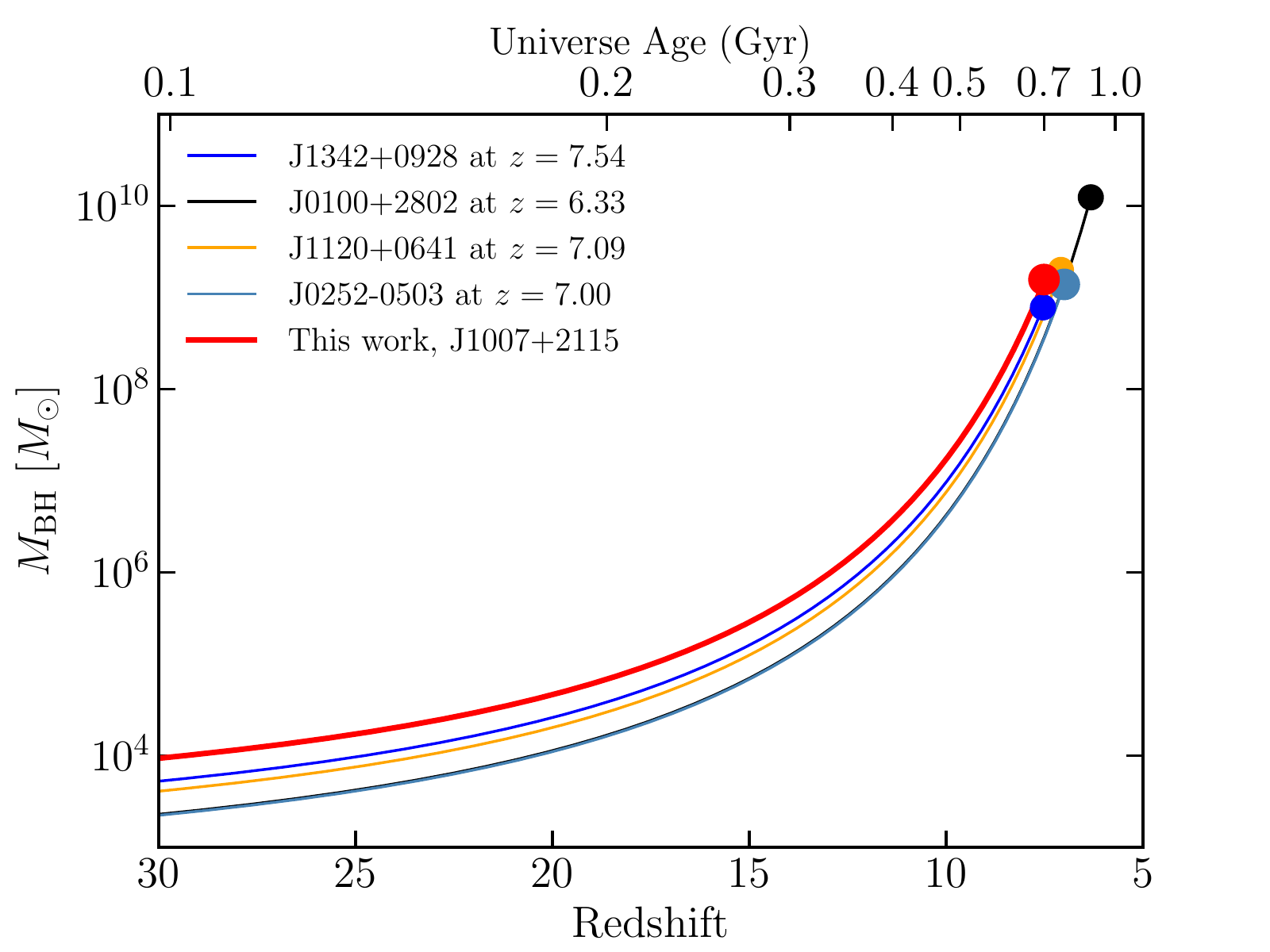} 
\caption{Black hole growth of J1007+2115, compared with those of quasars J1342+0928 at $z=7.54$ \citep{banados18}, J1120+0641 at $z=7.09$ \citep{mortlock11}, J0252--0503 at $z=7.00$ \citep{wang20}, and J0100+2802 at $z=6.33$ \citep{wu15}. The black hole growth is modeled as $M_{\rm BH}=M_{\rm seed}$exp[t/(0.05 Gyr)], assuming that the black holes accrete at the Eddington limit with a radiative efficiency of 0.1 since seed formation. The curves are normalized to the observed black hole mass and redshift of these quasars. J1007+2115 requires the most massive seed black hole under the same assumptions of black hole growth.}
\label{fig:bh}
\end{figure}

\begin{figure*}
\centering 
\epsscale{1.2}
\plotone{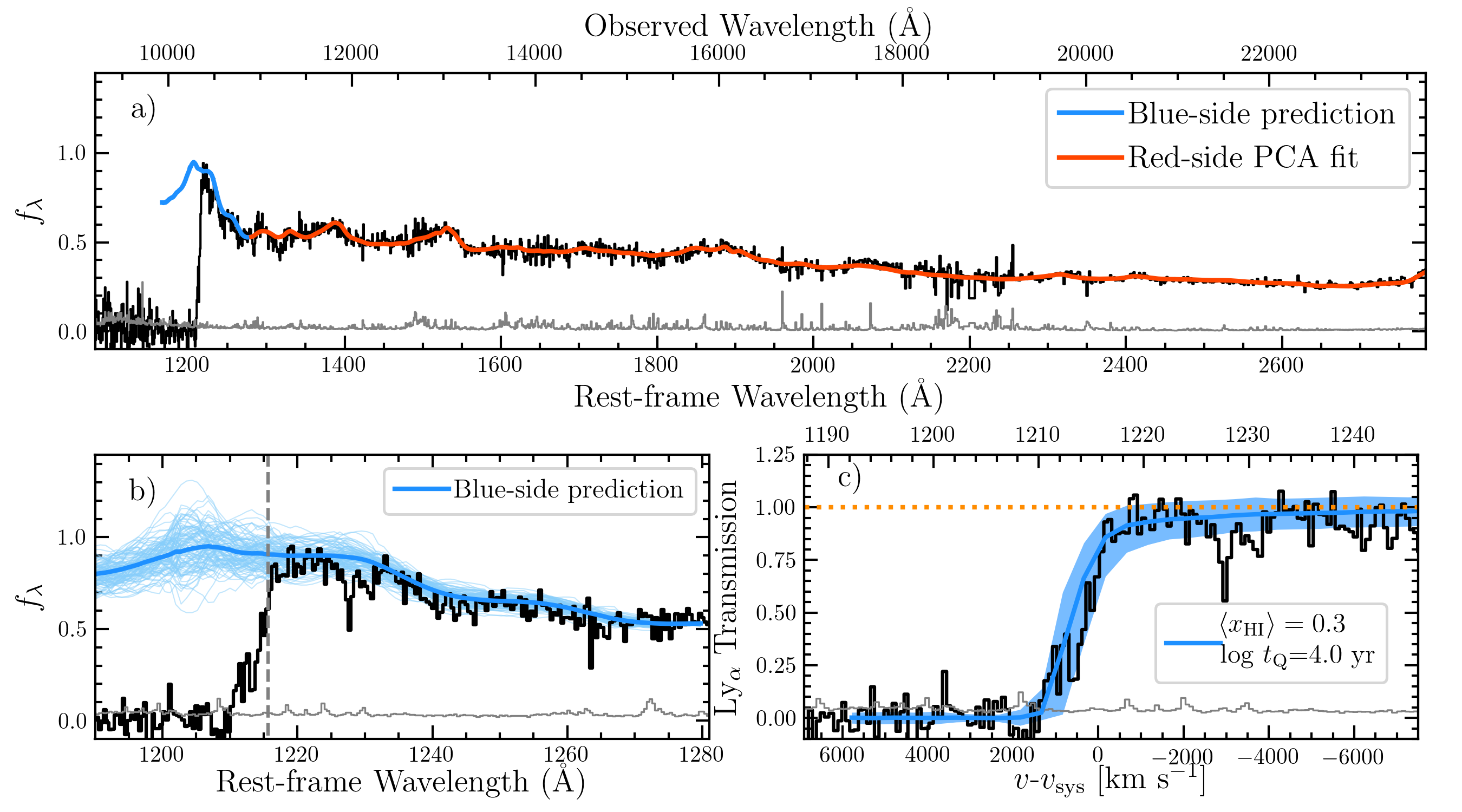}
\caption{{\bf a).} The intrinsic quasar spectrum from our PCA fit (red-side) and prediction (blue-side), compared with the observed spectrum in Figure \ref{fig:spec}. {\bf b).} The zoom-in Ly$\alpha$ region with 100 draws (thinner blue lines) from the covariant prediction error calibrated from the 1\% of most similar quasars in the PCA training sample. {\bf c).} The mock quasar transmission spectra with the volume-averaged neutral fraction $\langle x\rm_{HI}\rangle=0.3$ and quasar lifetime $t_{\rm Q}=10^{4.0}$ yr which are from the maximum pseudo-likelihood model. The solid blue line represents the median of mock spectra and the shaded region is the the 16th--84th percentile range.}
\label{fig:pca}
\end{figure*}

Observations of previously known luminous $z\gtrsim6.5$ quasars \citep[e.g.,][]{mortlock11, wu15, banados18} have already raised the question of how these early SMBHs grew in such a short time \citep[e.g.,][]{volonteri12, smith17, inayoshi19},
which probably requires massive seed black holes, as illustrated in Figure \ref{fig:bh}.
The black hole of J1007+2115, which is twice as massive as that of the
other $z=7.5$ quasar J1342+0928, further exacerbates this early SMBH
growth problem. To reach the observed SMBH mass at $z=7.5$,
a seed black hole with a mass of $\sim10^{4}\,(\mathrm{or}~3\times10^{5})\,M_{\rm
  \odot}$ would have to accrete continuously at the Eddington limit starting at $z=30\,(\mathrm{or}~15)$, 
assuming a radiative efficiency of 0.1 (see Figure \ref{fig:bh}). Under this
same set of fixed assumptions about black hole growth, J1007+2115 requires the most massive
seed black hole compared to any other known quasar.
This is consistent with the direct collapse black hole seed model rather than the Pop III stellar 
remnant seed model. Even with a massive seed black hole, Eddington accretion with a high duty cycle and
low radiative efficiency ($\sim 0.1$) is required. A lower mass seed would imply an even
higher accretion rate (e.g., hyper-Eddington accretion) or a lower radiative efficiency \citep{davies19}.
It has been suggested that maintaining super-Eddington accretion might be possible in specific
environments \citep{inayoshi16,smith17}, but whether or not this mode of rapid black hole growth is sustainable remains an important open question. 

\section{Constraint on the IGM Neutral Fraction from a Weak Damping Wing at $z=7.5$}

\begin{figure*}
\centering 
\epsscale{0.5}
\plotone{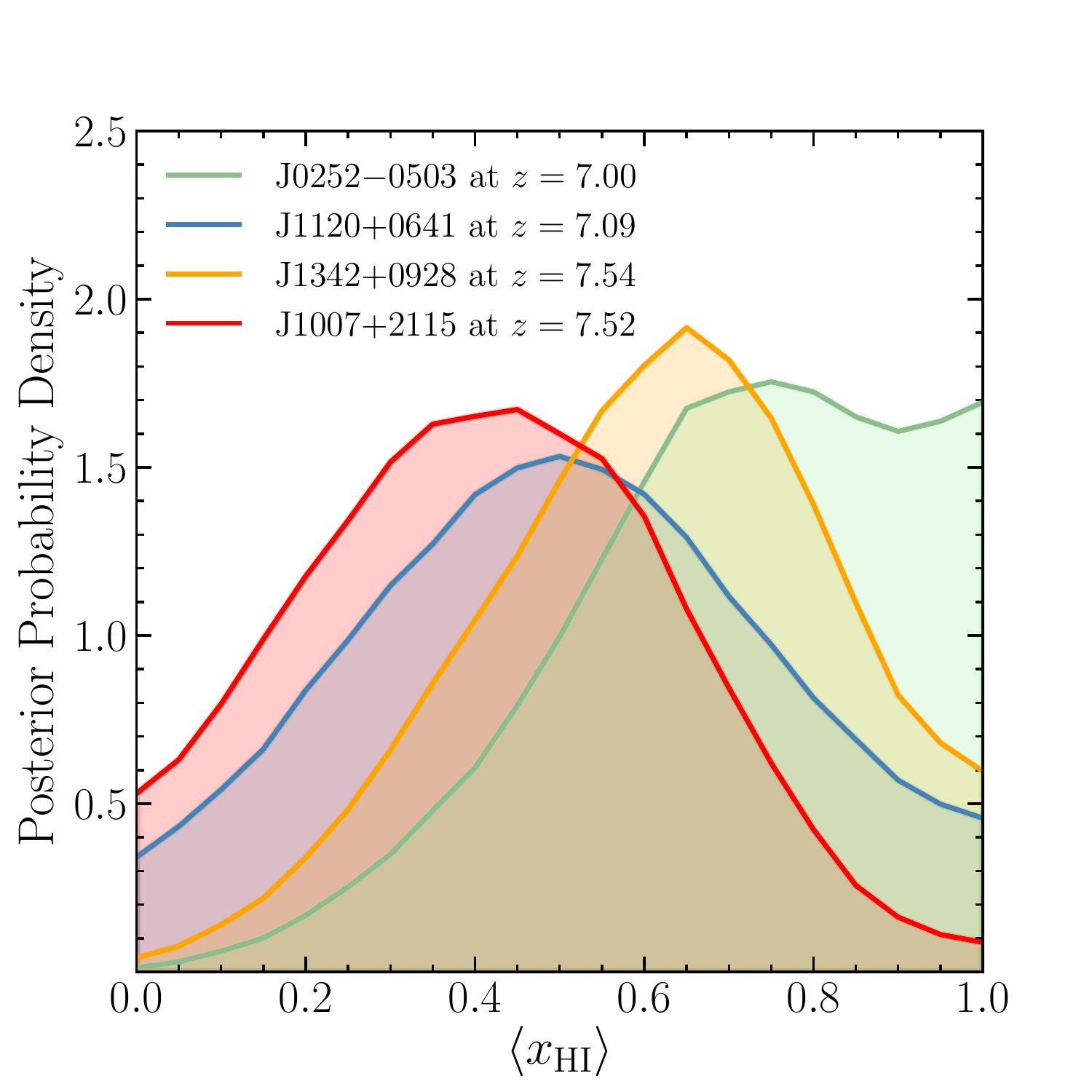}
\hspace{-15pt}
\epsscale{0.64}
\plotone{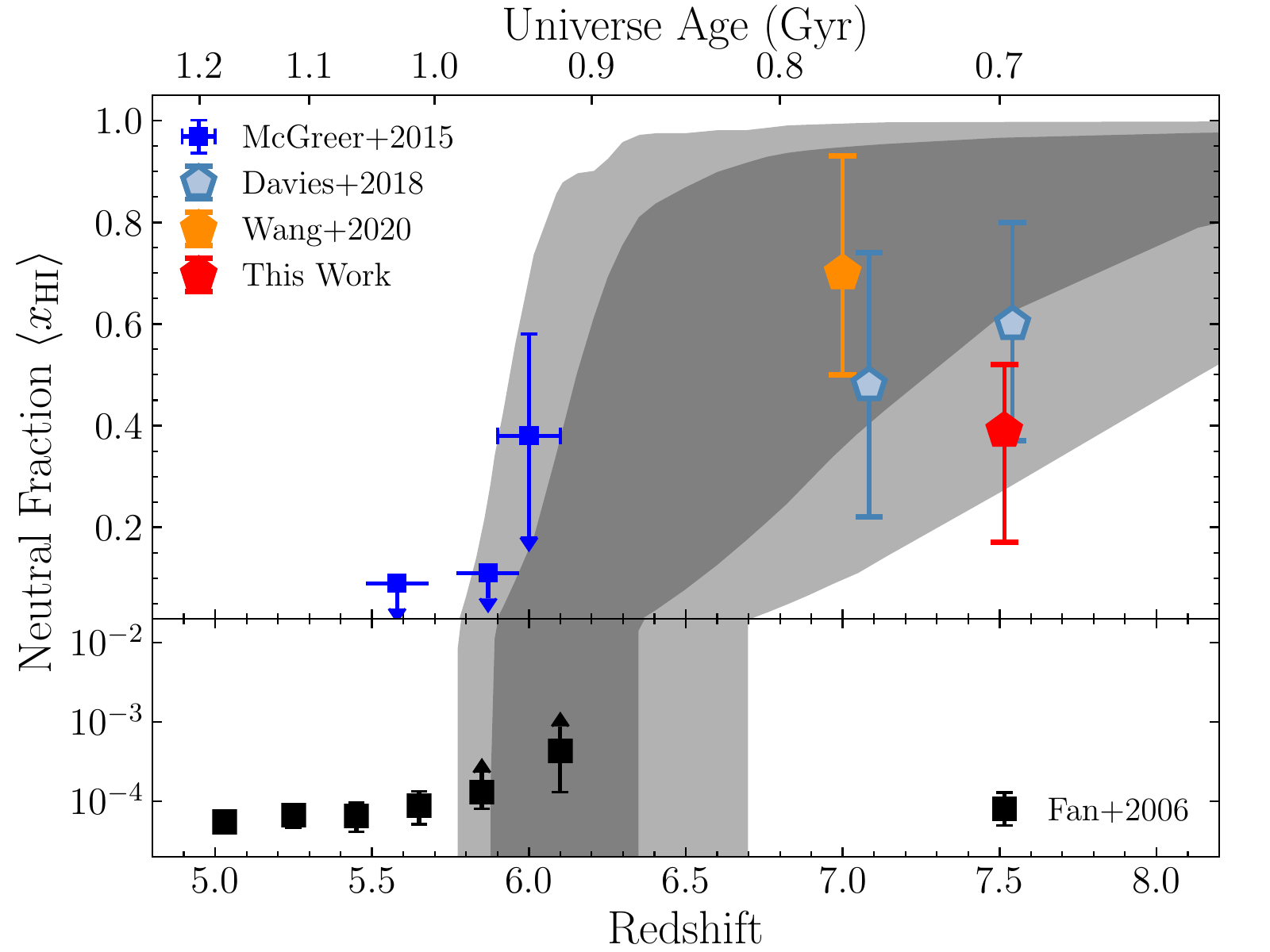} 
\caption{{\bf Left}: The posterior PDF of the volume-averaged neutral fraction $\langle x\rm_{HI}\rangle$ for J1007+2115, compared to $\langle x\rm_{HI}\rangle$ estimated from the other $z>7$ quasars that show a damping wing \citep{davies18a,wang20}. {\bf Right}: Constraints on the IGM neutral fraction derived from high redshift quasars through measurements of Ly$\alpha$ optical depth \citep[][black squares]{fan06}, dark gaps \citep[][blue squares]{mcgreer15}, and damping wings \citep[][blue and orange pentagons]{davies18a,wang20}. The new measurement for J1007+2115 is shown as the red filled pentagon. The dark and light grey shaded regions represent the 68\% and 95\% credible intervals from Planck observations \citep{planck18}. These quasar measurements indicate a rapidly changing phase from $z=7.5$ to $z=6$ with large scatter in the neutral fraction.}
\label{fig:xhi}
\end{figure*}

At $z>7$, the damping wing profile, detectable as absorption redward of the Ly$\alpha$ emission line caused by the highly neutral IGM, is one of the most promising tracers of the IGM neutral fraction. J1007+2115 provides us with a new sightline to estimate the IGM neutral fraction through damping wing analysis at a time deep into the reionization epoch. 

To estimate the IGM neutral fraction through damping wing analysis, 
we follow the procedures described in \cite{davies18a,davies18b},
which has also been used to analyze the spectra of three other
luminous $z\gtrsim 7$ quasars \citep{davies18a,wang20}. 
Briefly, we first model the quasar intrinsic continuum around the Ly$\alpha$ region using the principal component analysis (PCA) approach in \cite{davies18b}.
This approach predicts the intrinsic blue-side quasar spectrum (rest-frame 1175--1280 \AA) from the red-side spectrum (1280--2850 \AA) using a training sample of $\sim$13,000 quasar spectra from the SDSS/BOSS quasar catalog. 
We then apply the method from \cite{davies18a} to quantify the damping wing strength and estimate the volume-averaged neutral hydrogen fraction, $\langle x\rm_{HI}\rangle$. 
This method models the quasar transmission spectrum with a multi-scale hybrid model, which is a combination of the density, velocity, and temperature fields, large-scale semi-numerical reionization simulations around massive quasar-hosting halos (Davies \& Furlanetto in prep), and one-dimensional radiative transfer of ionizing photons emitted by the quasar \citep{davies16}. 
We construct realistic forward-modeled representations of quasar transmission spectra, accounting for the covariant intrinsic quasar continuum uncertainty. We then perform Bayesian parameter inference on the mock spectra to recover the joint posterior probability distribution functions (PDF) of $\langle x\rm_{HI}\rangle$ and $\log t_\mathrm{Q}$ from the observed spectrum. In the Bayesian inference, the likelihood is computed from maximum pseudo-likelihood model parameters and the pseudo-likelihood is defined as the product of individual flux PDFs of 500 km/s binned pixels, equivalent to the likelihood function of the binned transmission spectrum in the absence of correlations between pixels (see more details in \citealt{davies18a}).

To measure $\langle x\rm_{HI}\rangle$, we set a broad $t_{\rm Q}$ range of $10^3\,{\rm yr}<t_{\rm Q}<10^8\,{\rm yr}$ with a flat log-uniform prior, and compute the posterior PDF of $\langle x\rm_{HI}\rangle$ by marginalizing over quasar lifetime. 
As shown in Figure \ref{fig:xhi}, from the posterior PDF, we can estimate $\langle x\rm_{HI}\rangle$ and its 68\% confidence interval as $\langle x\rm_{HI}\rangle=0.39^{+0.22}_{-0.13}$, which is consistent with the maximum pseudo-likelihood model parameters shown in Figure \ref{fig:pca}.
To avoid possible contamination from any intervening damped Ly$\alpha$ absorber, we search for associated metal absorption. No such absorption has been found in our current spectrum. We conclude that the neutral IGM should be responsible for the damping wing features of J1007+2115. If any potential damped Ly$\alpha$ absorber plays a role in generating the damping wing feature, the IGM neutral fraction will be even lower.

The detection of damping wing signatures in two $z>7$ quasar spectra has previously provided strong evidence for a significantly neutral Universe at $z\gtrsim7$ \citep[e.g.,][]{mortlock11,banados18, davies18a}. Specifically, neutral gas fractions of $\langle x\rm_{HI}\rangle\sim0.48$ at $z=7.09$ and $\langle x\rm_{HI}\rangle\sim0.60$ at $z=7.54$ have been reported \citep{davies18a}. Recent analysis of the damping wing feature of the quasar J0252--0503 at $z=7.0$ \citep{wang20} also suggests a highly neutral IGM with $\langle x\rm_{HI}\rangle=0.7$. All of these measurements are based on the same methodology used in this work. We compare our result with these estimates, as shown in Figure \ref{fig:xhi}.
It is evident that the damping wing absorption is much weaker in J1007+2115 compared to that in the other three $z>7$ quasars. At the resonant Ly$\alpha$ wavelength, the observed spectrum of J1007+2115 does not deviate from the blue-side prediction based on red-side PCA reconstruction. 
This result is to be compared with J0252--0503 \citep{wang20} where the observed spectrum is $\sim 40$\% lower than the prediction without damping wing absorption. 
The $\langle x\rm_{HI}\rangle$ estimated from J1007+2115 at $z=7.54$ is lower than the measurements from all of the other three sightlines. 
Studies of the Ly$\alpha$ emission from $z>6$ galaxies have suggested neutral fractions of $\langle x\rm_{HI}\rangle=0.59^{+0.11}_{-0.15}$ at $z\sim7$ and $\langle x\rm_{HI}\rangle>0.76$ at $z\sim8$ \citep{mason18,mason19}. The sightline of J1007+2115 is thus a $2\sigma$ outlier, compared to the previous results.
Although it is difficult to draw solid conclusions because of the large uncertainties (and the broad PDF) on the value of $\langle x\rm_{HI}\rangle$, the much weaker damping wing seen in J1007+2115's spectrum indicates a significant scatter of the IGM neutral fraction in the redshift range $z=7.5$ to $z=7.0$, which can be interpreted as observational evidence of patchy reionization.

\section{Summary}
We report the discovery of a new quasar J1007+2115 with a \cii-based redshift of $z=7.5149\pm0.0004$, selected with DECaLS, PS1, UHS, and WISE photometry and observed with the Gemini, Magellan, Keck, and ALMA telescopes. 
The \cii\ and dust continuum emission from the quasar host galaxy are well detected, and imply a SFR$_\mathrm{[CII]}\sim80-520$\,$M_\sun$\,yr$^{-1}$. It is only the second quasar known at such high redshift and thus provides a valuable new data point for early SMBH and reionization history studies.  

By fitting the NIR spectrum, we derive $M_{\rm BH}=(1.5\pm0.2)\times10^{9}$\,$M_{\odot}$ and an Eddington ratio of $L_{\rm bol}/L_{\rm Edd}=1.06\pm0.2$ using the broad \mgii\ emission line.
The black hole in J1007+2115 is twice as massive as that of J1342+0928 at a very similar redshift of $z=7.54$, and thus places the strongest constraint to the early SMBH growth, requiring a seed black hole with a mass of $\sim10^{4}\,(3\times10^{5})\,M_{\rm \odot}$ at $z=30\,(15)$. 
Through damping wing modeling of the quasar spectrum, we estimate the volume-averaged neutral fraction to be $\langle x\rm_{HI}\rangle=0.39^{+0.22}_{-0.13}$ at $z=7.5$. Together with three previous measurements from quasar damping wing analyses, our new result indicates a large scatter of the IGM neutral fraction from $z=7.5$ to $z=7.0$, indicative of a patchy reionization process. 

\begin{deluxetable}{l l}
\setlength{\tabcolsep}{15pt}
\tabletypesize{\scriptsize}
\tablecaption{Photometric Properties and Derived Parameters of J1007+2115.}
\tablewidth{0pt}
\startdata
\\
R.A. (J2000) & 10:07:58.26 \\
Decl. (J2000) & +21:15:29.20 \\
$z_{\rm [CII]}$ & 7.5149$\pm$0.0004 \\
$m_{1450}$ & 20.43$\pm$0.07\\
$M_{1450}$ & --26.66 $\pm$0.07 \\
$f_{\lambda,z \rm PS1}$\tablenotemark{a} & 1.3$\times$$10^{-18}$ erg s$^{-1}$ cm$^{-2}$ \AA$^{-1}$  \\
$f_{\lambda,y \rm PS1}$\tablenotemark{a} & 2.1$\times$$10^{-18}$ erg s$^{-1}$ cm$^{-2}$ \AA$^{-1}$  \\
$f_{\lambda,z \rm DECaLS}$\tablenotemark{a} & 6.2$\times$$10^{-19}$ erg s$^{-1}$ cm$^{-2}$ \AA$^{-1}$ \\
$Y$   & 21.30$\pm$0.13 \\
$J$   & 20.20$\pm$0.18\\
$H$   & 20.00$\pm$0.07\\
$K$   & 19.75$\pm$0.08\\
$W1$   & 19.56$\pm$0.11\\ 
$W2$   & 19.44$\pm$0.20\\
\hline
$z_{\rm MgII}$ & 7.494$\pm$0.001\\
$z_{\rm CIV}$ & 7.403$\pm$0.01\\
$\alpha_\lambda$ & --1.14$\pm$0.01\\
$\rm \Delta v_{MgII-[CII]}$ & -736$\pm$35 km s$^{-1}$\\
$\rm \Delta v_{CIV-MgII}$ & -3220$\pm$362 km s$^{-1}$\\
FWHM$\rm_{MgII}$ & 3247$\pm$188 km s$^{-1}$\\
FWHM$\rm_{CIV}$ & 6821$\pm$2055 km s$^{-1}$\\
$\lambda L_{\rm 3000\text{\normalfont\AA}}$ & (3.8$\pm$0.2)$\times$$10^{46}$ erg s$^{-1}$\\
$L_{\rm bol}$ & (1.9$\pm$0.1)$\times$$10^{47}$ erg s$^{-1}$\\
$M_{\rm BH}$ & (1.5$\pm$0.2)$\times$10$^{9}$ $M_{\odot}$\\
$L_{\rm bol}/L_{\rm Edd}$ & 1.06$\pm$0.2\\
\hline
$F_\mathrm{[CII]}$  &   1.2$\pm$0.1  Jy\,\kms\\
FWHM$_\mathrm{[CII]}$ & 331.3$\pm$31.6 \kms \\
$L_\mathrm{[CII]}$ & (1.5$\pm$0.2)$\times$10$^{9}$ $L_{\odot}$\\
S$_{231.2\ \rm GHz}$ &  1.2$\pm$0.03 mJy \\
SFR$_\mathrm{[CII]}$ & 80 -- 520 $M_\sun$\,yr$^{-1}$\\
SFR$_\mathrm{TIR}$ & 700 $M_\sun$\,yr$^{-1}$\\
\enddata
\tablenotetext{a}{They are 3-$\sigma$ flux limits in PS1 $z$, PS1 $y$, and DECaLS $z$ bands, from our forced photometry with 3 arcsec aperture diameter.}
\label{tab:table1}
\end{deluxetable}

\acknowledgments
Thanks to Dave Osip for approving the request of FIRE spectropsopy which is important to the confirmation of this quasar.
J. Yang, X. Fan and M. Yue acknowledge the supports from the NASA ADAP Grant NNX17AF28G.
F. Wang thanks the support provided by NASA through the NASA Hubble Fellowship grant \#HST-HF2-51448.001-A awarded by the Space Telescope Science Institute, which is operated by the Association of Universities for Research in Astronomy, Incorporated, under NASA contract NAS5-26555.
L. Jiang and X.-B. Wu thank the support from the National Key R\&D Program of China (2016YFA0400703) and the National Science Foundation of China (11533001 \& 11721303).
Research by A.J.B. is supported by NSF grant AST-1907290.

Some of the data presented in this paper were obtained at the W.M. Keck Observatory, which is operated as a scientific partnership among the California Institute of Technology, the University of California and the National Aeronautics and Space Administration. The Observatory was made possible by the generous financial support of the W. M. Keck Foundation.
The authors wish to recognize and acknowledge the very significant cultural role and reverence that the summit of Maunakea has always had within the indigenous Hawaiian community.  We are most fortunate to have the opportunity to conduct observations from this mountain.
This research is based in part on observations obtained at the Gemini Observatory (GN-2019B-Q-135, GN-2019A-DD-109), which is operated by the Association of Universities for Research in Astronomy, Inc., under a cooperative agreement with the NSF on behalf of the Gemini partnership: the National Science Foundation (United States), National Research Council (Canada), CONICYT (Chile), Ministerio de Ciencia, Tecnolog\'{i}a e Innovaci\'{o}n Productiva (Argentina), Minist\'{e}rio da Ci\^{e}ncia, Tecnologia e Inova\c{c}\~{a}o (Brazil), and Korea Astronomy and Space Science Institute (Republic of Korea).
This paper makes use of the following ALMA data: ADS/JAO.ALMA\#2019.1.01025.S. ALMA is a partnership of ESO (representing its member states), NSF (USA) and NINS (Japan), together with NRC (Canada), MOST and ASIAA (Taiwan), and KASI (Republic of Korea), in cooperation with the Republic of Chile. The Joint ALMA Observatory is operated by ESO, AUI/NRAO and NAOJ. The National Radio Astronomy Observatory is a facility of the National Science Foundation operated under cooperative agreement by Associated Universities, Inc.
UKIRT is owned by the University of Hawaii (UH) and operated by the UH Institute for Astronomy; operations are enabled through the cooperation of the East Asian Observatory.
This paper includes data gathered with the 6.5 meter Magellan Telescopes located at Las Campanas Observatory, Chile.
We acknowledge the use of the PypeIt data reduction package.

Thanks to the Leo Ola program for naming this quasar in Hawaiian language.
The name P\={o}niu\={a}`ena is a result of the Leo Ola program offered through the University of Hawai`i at Hilo's Ka Haka `Ula o Ke`elik\={o}lani, College of Hawaiian Language and the `Imiloa Astronomy Center. The efforts undertaken through Leo Ola help to create a pathway where language and culture are at the core of modern scientific practices, melding indigenous culture and science locally, nationally and worldwide.


\facilities{Gemini(GMOS), Keck(NIRES), Magellan(FIRE), UKIRT(WFCam), ALMA}

\software{PypeIt \citep{prochaska20}}

\end{document}